\documentclass{elsart}
\usepackage{natbib}
\usepackage{graphics}
\usepackage{amssymb}
\usepackage{aas_macros}

\begin{document}
\begin{frontmatter}
\title{Metric Radio Bursts and Fine Structures Observed on 17 January 2005}
\author[UOA]{C. Bouratzis},
\author[UOA]{P. Preka-Papadema\corauthref{PPP}},
\corauth[PPP]{P. Preka-Papadema} \ead{ppreka@phys.uoa.gr}
\author[UOA]{X. Moussas},
\author[UOI]{C. Alissandrakis}
\and
\author[UOA]{A. Hillaris}

\address[UOA]{Department of Physics, University of Athens, GR-15784 Athens, Greece}
\address[UOI]{Department of Physics, University of Ioannina, 45110 Ioannina, Greece}

\begin{abstract}
A complex radio event was observed on January 17, 2005 with the radio-spectrograph ARTEMIS-IV,
operating at Thermopylae, Greece; it was associated with an X3.8 SXR flare and two fast Halo
CMEs in close succession. We present ARTEMIS--IV dynamic spectra of this event;
the high time resolution (1/100 sec)  of the data in the 450--270 MHz range, makes
possible the detection and analysis of the fine structure which this major radio event exhibits.
The fine structure was found to match, almost, the comprehensive Ondrejov Catalogue
which it refers to the spectral range 0.8--2 GHz, yet seems to produce
similar fine structure with the metric range.
\end{abstract}

\end{frontmatter}
\parindent=0.5 cm
\section{INTRODUCTION}\label{Introduction}

Radio emission at metric and longer waves trace disturbances,
mainly electron beams and shock waves, formed in the process of energy release and
magnetic restructuring of the corona and propagating from the low corona to interplanetary
space. The fine structures, on the other hand,
including drifting pulsation structures, may be used as powerful diagnostics
of the loop evolution of solar flares.

The period 14--20 January 2005 was one of intense activity originating in
active regions 720 and 718; while in the visible hemisphere of the sun,
they produced 5 X--class and 17 M--class flares, an overview is presented in ~\cite{Bouratzis06}.

January the 17th is characterized by an X3.8 SXR flare from 06:59 UT to 10:07 UT (maximum at 09:52 UT) and two fast Halo
CMEs within a forty minute interval.
The corresponding radio event included an extended broadband continuum with rich fine structure;
this fine structure is examined in this report.

\section{Observations} \label{obs}
\subsection{Instrumentation}
The Artemis IV\footnote{Appareil de Routine pour le Traitement et l'
Enregistrement Magnetique de l' Information Spectral} solar radio-spectrograph operating at Thermopylae since 1996
\citep{Caroubalos01, Kontogeorgos06} consists of a 7-m parabolic antenna covering the metric range, to which a dipole
antenna was added recently in order to cover the decametric range. Two
receivers operate in parallel, a sweep frequency analyzer (ASG) covering the
650-20 MHz range in 630 data channels with a cadence of 10 samples/sec and a
high sensitivity multi-channel acousto-optical analyzer (SAO), which covers
the 270-450 MHz range in 128 channels with a high time resolution of 100
samples/sec.

Events observed with the instrument have been described, e.g. by
\citet{Caroubalos04,Caroubalos01B}, \citet{Kontogeorgos,Kontogeorgos08},
\citet{Bouratzis06}, \citet{Petoussis06}, cf. also \citet{Caroubalos06} for a brief review.

The broad band, medium time resolution recordings of the ASG are used for the detection and analysis of
radio emission from the base of the corona to two $R_{SUN}$, while the narrow band, high time
resolution SAO recordings are mostly used in the analysis of the fine temporal and spectral structures.

\subsection{The event of January 17, 2005--Overview}\label{Overview}
\begin{figure}[]
\begin{center}
 \includegraphics{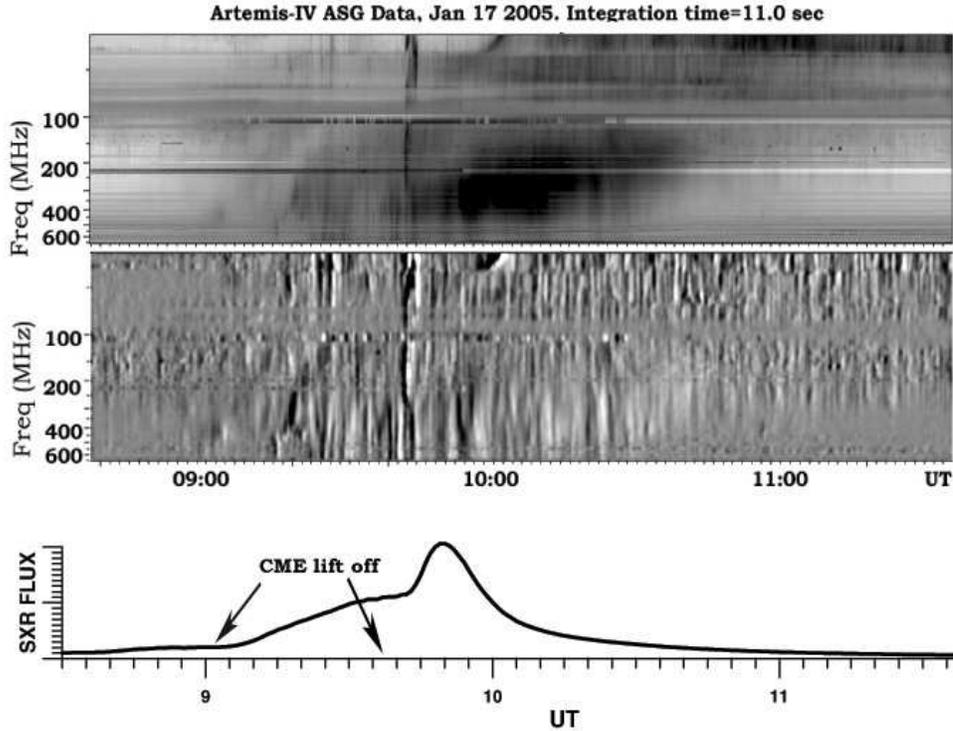}
 \caption{ARTEMIS IV Dynamic Spectrum (08:40-11:30 UT). UPPER PANEL: ASG Spectrum,
MIDDLE PANEL: ASG Differential Spectrum, LOWER PANEL: GOES SXR flux (arbitrary units);
the two CME lift--offs are marked on the time axis.}
   \label{05117_01}
\end{center}
\end{figure}
\begin{figure}[]
\begin{center}
 \includegraphics{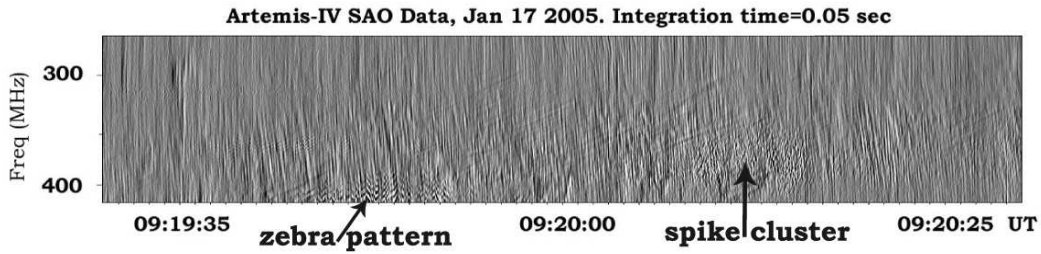}
 \caption{ARTEMIS IV SAO Differential Spectrum: Zebra pattern (09:19:40-09:19:52 UT) and a spike cluster
(09:20:08-09:20:15) on a background of fiber bursts \& pulsations (09:19:30-09:20:30 UT);
the pulsations with the fibers cover the observation period but they appear more pronounced in
the 09:35-09:55 UT interval.}
   \label{05117_01FS}
\end{center}
\end{figure}
\begin{figure}[]
\begin{center}
 \includegraphics{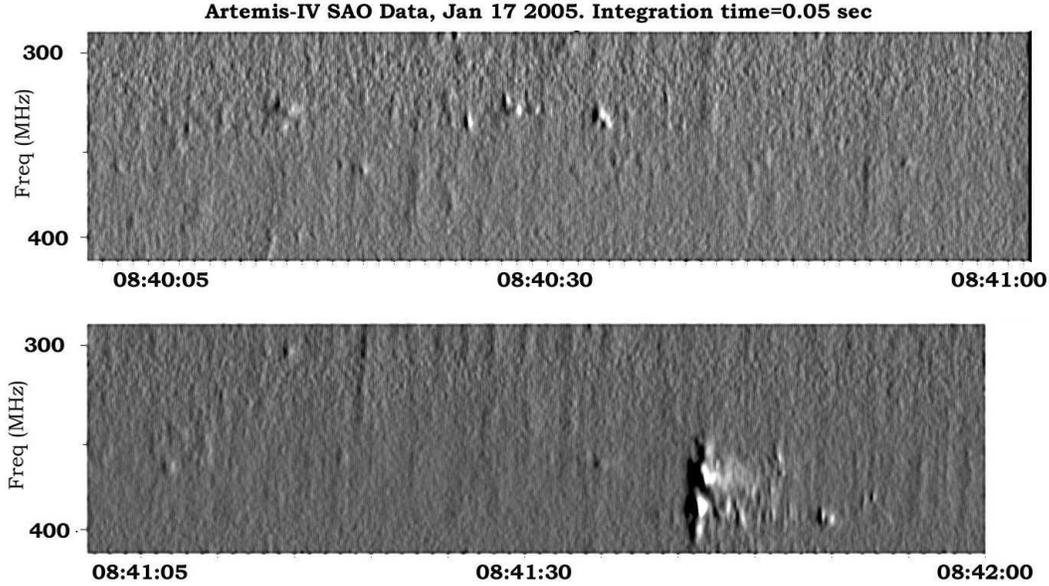}
 \caption{ARTEMIS IV Differential Spectra of Fine Structures embedded in the Type IV
Continuum. UPPER PANEL: Spikes, LOWER PANEL: Narrow Band Type III(U) Bursts.}
   \label{FS1}
\end{center}
\end{figure}
\begin{figure}[]
\begin{center}
 \includegraphics{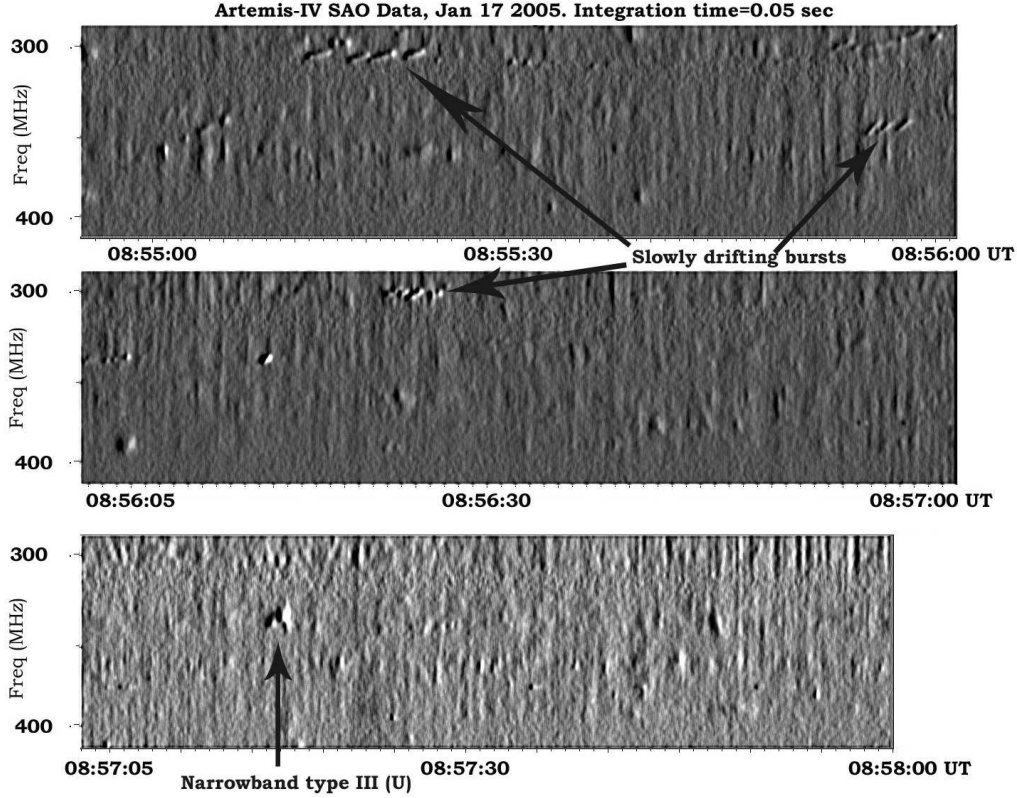}
 \caption{ARTEMIS IV Differential Spectra of Fine Structures embedded in the Type IV
Continuum. FIRST TWO PANELS: Narrow Band Slowly Drifting Bursts LOWER PANEL: Narrow Band Type III(U).}
   \label{FS1B}
\end{center}
\end{figure}
\begin{figure}[]
\begin{center}
 \includegraphics{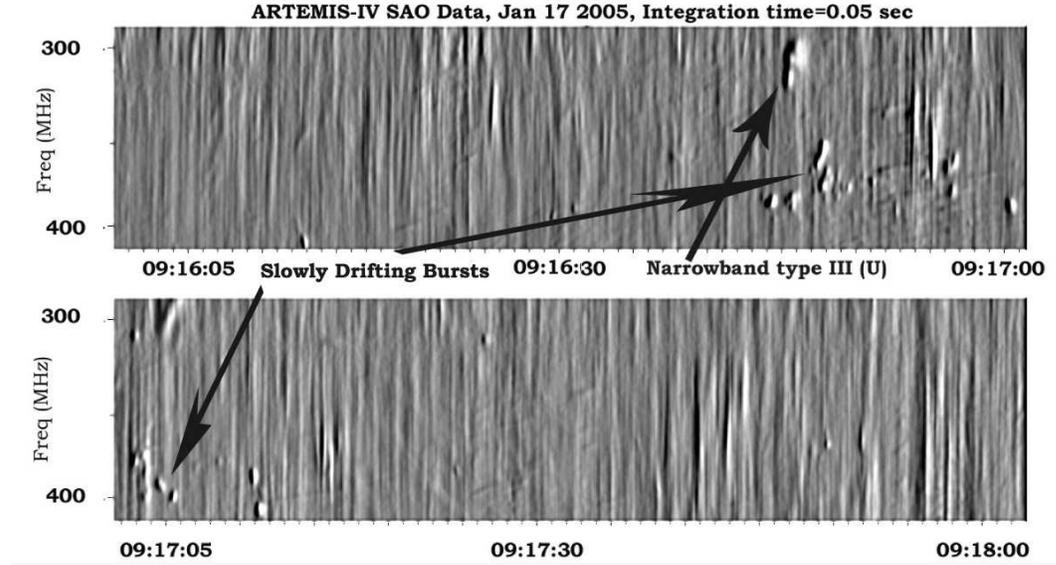}
 \caption{ARTEMIS IV Differential Spectra of Fine Structures embedded in the Type IV
Continuum; narrowband bursts within groups fiber bursts \& pulsating structures:
UPPER PANEL: Narrowband Type III(U),  Narrow Band Slowly Drifting Bursts \& spikes;
LOWER PANEL: Narrow Band Slowly Drifting Burst \& spikes.}
   \label{FS1C}
\end{center}
\end{figure}
\begin{figure}[]
\begin{center}
 \includegraphics{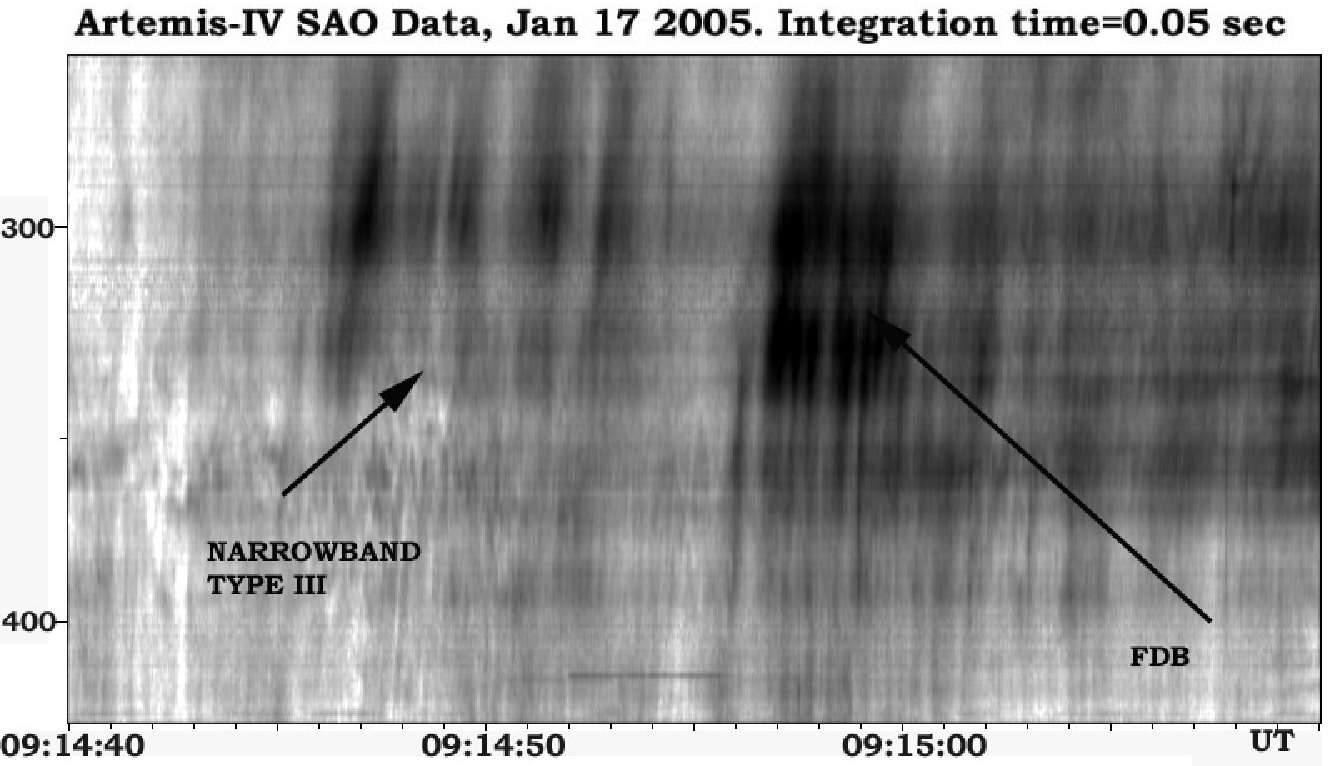}
 \includegraphics{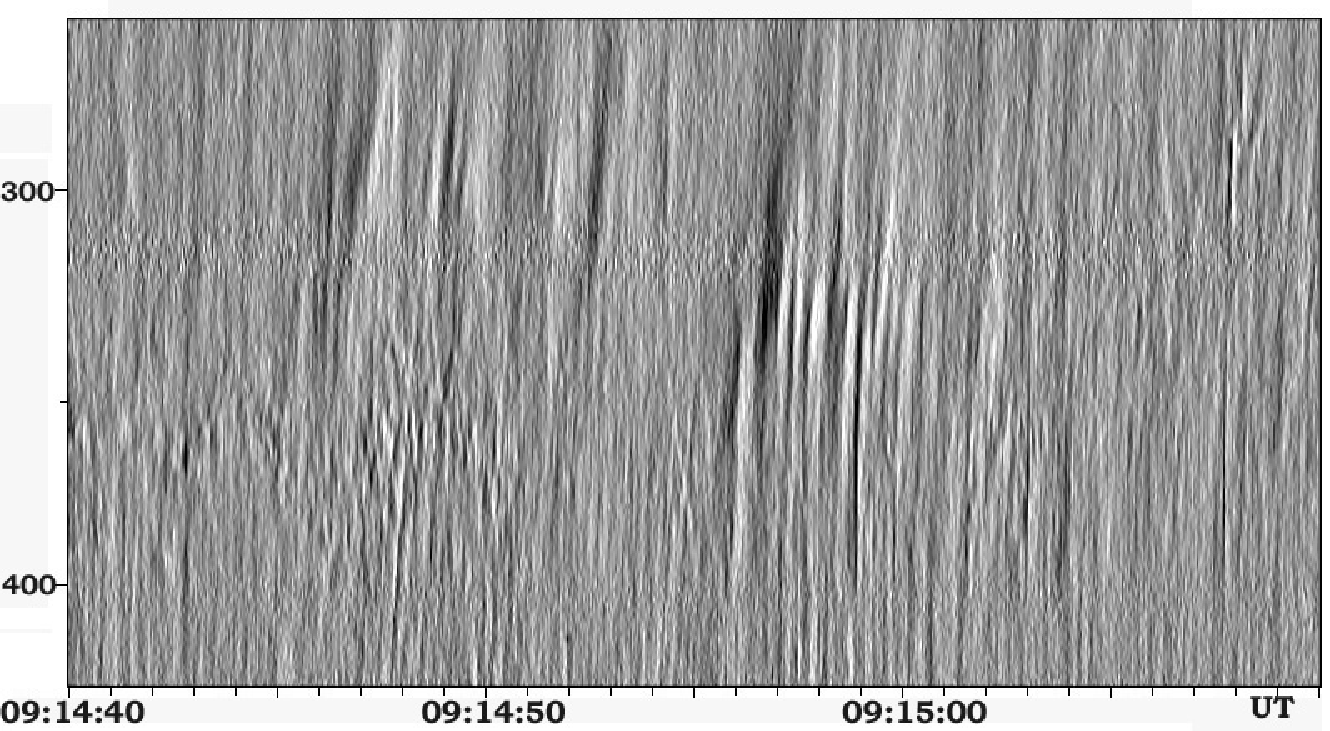}
 \caption{ARTEMIS IV Spectra of Fast Drift Bursts (FDB), 09:14:55-09:15:00 UT,
preceded by narrowband type III bursts. 
UPPER PANEL: Intensity Spectrum LOWER PANEL: Differential Spectrum.}
   \label{FDB}
\end{center}
\end{figure}
\begin{figure}[]
\begin{center}
 \includegraphics{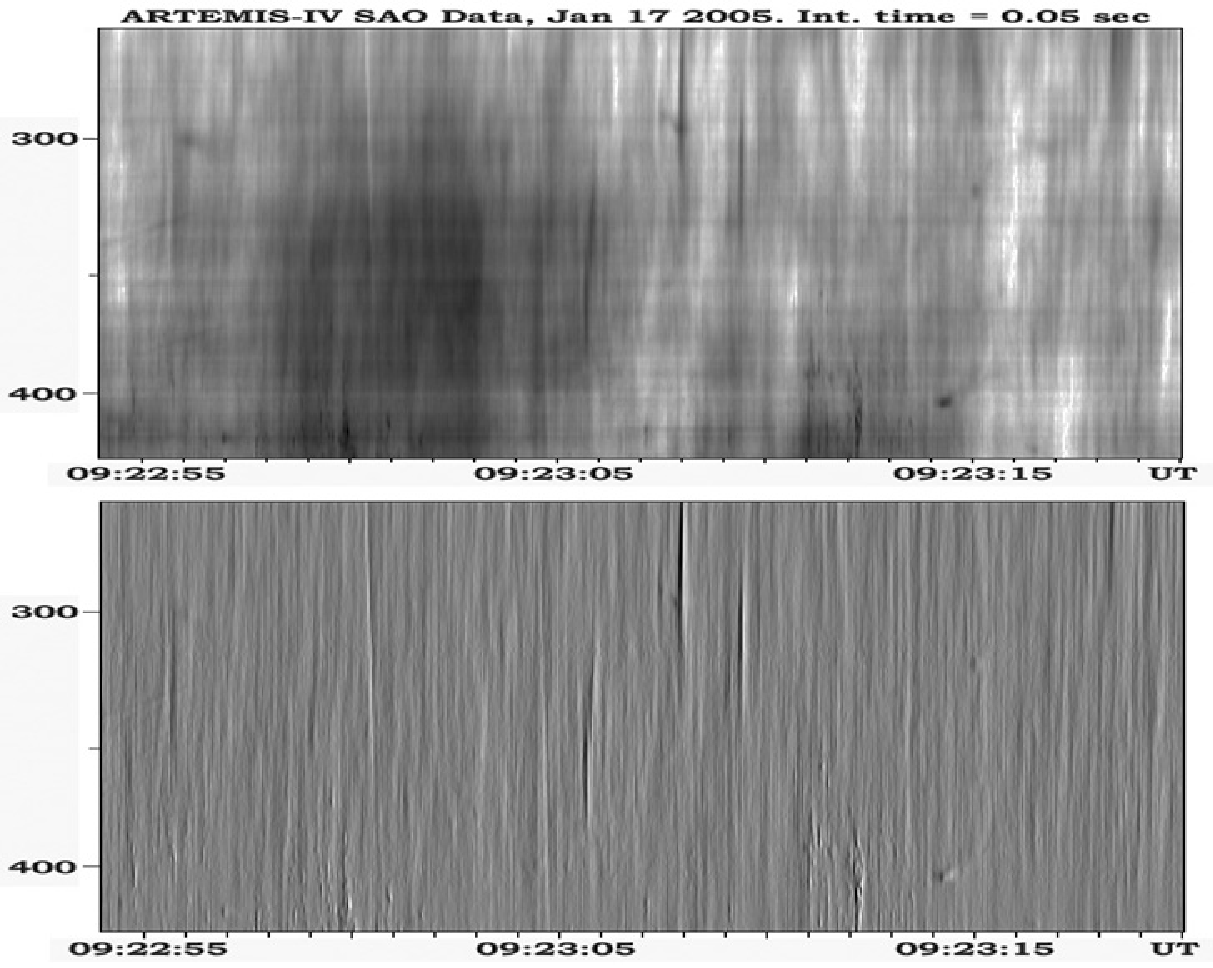}
 \caption{ARTEMIS IV Spectra of Isolated Pulsating Structures (IPS), 09:22:55-09:23:05 UT.
UPPER PANEL: Intensity Spectrum LOWER PANEL: Differential Spectrum.}
   \label{IPS}
\end{center}
\end{figure}
Two groups of type III bursts and a very extensive type IV continuum with
rich fine structure characterize the ARTEMIS--IV dynamic spectrum.
The high frequency type IV emission starts at 08:53 UT, covers the entire 650-20MHz ARTEMIS--IV
spectral range (Figure \ref{05117_01}, upper \& middle panels)
and continues well after 15:00; it was associated with an SXR flare and two fast
Halo CMEs (CME$_1$ \& CME$_2$ henceforward) in close succession.

The GOES records\footnote{http//www.sel.noaa.gov/ftpmenu/indices}
report an X3.8 SXR flare from 06:59 UT to 10:07 UT, with maximum
at 09:52 UT; this is well associated with  the brightening of sheared S-shaped loops from
the  EIT images. The SXR light curve (Figure \ref{05117_01}, lower panel) exhibits an, initially,
slow rising phase which changes into a much faster rising a little before the peak flux is reached; it thus
appears on the time--SXR flux diagram as a two stage process. The CME data from the LASCO lists on
line\footnote{http://cdaw.gsfc.nasa.gov/CME list}
\citep{Yashiro} indicate that each of the stages coincides with the, estimated,
lift off time of CME$_1$ \& CME$_2$ respectively; its also well associated with the high frequency onset of the
two type III groups mentioned in the beginning of this subsection.
The \emph{halo} CME$_1$ was first recorded by LASCO at 09:30:05 UT.
Backward extrapolation indicates that it was launched around
09:00:47 UT. CME$_2$ was first recorded by LASCO at 09:54 UT; it was launched around
09:38:25 UT and was found to overtake CME$_1$ at about 12:45 UT at a height of approximate 37 solar radii.

\subsection{Fine Structure}\label{FS}
The high sensitivity and time resolution of the SAO facilitated an examination on fine structure
embedded in the Type--IV continua within the studied period. In our analysis, the continuum background is
removed by the use  of high--pass filtering on the dynamic spectra (differential spectra in this case).

As fine structure is characterized by a large variety in period, bandwidth, amplitude, temporal
and spatial signatures, a morphological taxonomy scheme
based on Ondrejov Radiospectrograph recordings in the 0.8--2.0 GHz range was introduced
(\citet{Jiricka01,Jiricka02} also \citet{Meszarosova05B});
the established classification criteria are used throughout this report.

We present certain examples of fine structures recorded by the
ARTEMIS--IV/SAO in the 450-270 MHz frequency range; this range
corresponds to ambient plasma densities which are typical of the
base of the corona ($\approx 10^9$~$cm^{-3}$, (cf. for example \citeauthor{Mann99}, \citeyear{Mann99}).
The fine structures of our data set, are divided according to the above mentioned Ondrejov
classification scheme and described in the following paragraphs
(cf. also figures \ref{05117_01FS}, \ref{FS1}, \ref{FS1B} \& \ref{FS1C}).

\subsubsection{Broadband pulsations \& Fibers}
The broadband pulsations appear for the duration of the type IV continuum; they are, for
same period, associated with fibers; these structures intensify within the rise phase of the
SXR, which in turn, coincides with the extrapolated liftoff of CME$_1$ \& CME$_2$. Follows
a closer examination:
\begin{itemize}
\item{Radio pulsations are series of pulses with bandwidth $> 200 MHz$ and total duration $> 10 s$;
on the ARTEMIS-IV recordings they persisted
for the duration of the type IV continuum. Some, with a slow global frequency drift, were of
the \textit{Drifting Pulsation Structures} (DPS) subcategory. In our recordings
the pulsations bandwidth exceeded the SAO frequency range, however from the ASG dynamic spectrum
we observe a drift of the pulsating continuum towards the lower frequencies, following the rise of CME$_2$.
Three physical mechanisms were proposed as regards the source of this type of structure, (cf \citet{Nindos07}
for a review):
\begin{itemize}
\item{Modulation of radio emission by MHD oscillations in a coronal loop}
\item{Non-linear oscillating
systems (wave-wave or  wave particle interactions) where the pulsating structure corresponds to their limit cycle}
\item{Quasi-periodic injection of electron populations from acceleration episodes within large scale current sheets.}
\end{itemize}
Combined radiospectrograph, radio and SXR imaging and HXR observations \citep{Kliem00,Khan02,Karlicky02}
favor the last mechanism; furthermore they identify the sources of Drifting Pulsation Structures
with plasmon ejections.}
\item{Isolated Broadband Pulses: Pulsating Structures but with duration $\approx 10 s$.}
\item{Fast Drifting Bursts: Short-lasting and fast drifting bursts with frequency
drift $>100MHz/s$; similar to the isolated broadband pulses, except for the frequency drift.}
\item{Fibers or Intermediate Drift Bursts: Fine structure Bursts with the frequency drift $\approx 100 MHz/s$;
they often  exhibit nearly regular repetition. On our recordings they coincide with broadband pulsations and they also
cover the duration of the type IV continuum. They are usually interpreted as the radio signature of whistler waves
coalescence with Langmuir waves in magnetic loops; the exciter is thought to be an unstable
distribution of nonthermal electrons. (cf. \citet{Nindos07} and references within).}
\end{itemize}

\subsubsection{Zebra patterns:}
The zebra structures are characterized by several emission lines,
which maintain nearly regular distance to their neighbors (figure \ref{05117_01FS}).
The zebras from our data set are associated with
pulsations and fibers and cover almost the same period with them.
They appear, however, more pronounced in the 09:35--09:55 UT
interval; this interval includes the $CME_2$ estimated launch and the rise phase of the
SXR flare. Zebra patterns were explained as the result electrostatic upper-hybrid waves at conditions of the double
plasma resonance where  the local upper hybrid frequency equals a multiple of the local gyrofrequency 
($\omega _{UH}  = \sqrt {\omega _e^2  + \omega _{Be}^2 }  = s \cdot \omega _{Be}$)
(cf. for example \citet{Chernov06, Nindos07} and references therein).  
The upper hybrid waves are excited by electron beam with loss-cone distribution \citep{Kuznetsov07}.
\subsubsection{Narrowband Structures}
The narrowband activity (figures \ref{FS1}, \ref{FS1B} \& \ref{FS1C}), including Spikes, Narrow
Band Type III \& III(U) bursts as well as Slowly Drifting Structures, is rather intermittent.
A large group of spikes appears at about 09:20 UT; this coincides, in time, with the rising of the first stage
of the SXR and the start of the first type III group. Three types of Narrowband Structures were recorded:
\begin{itemize}
\item{Narrow Band Spikes are very short ($\approx 0.1 s$) narrowband ($\approx 50 MHz$) bursts which
usually appear in dense clusters. An example of such a cluster appears on figure \ref{05117_01FS}.
The models proposed for the spike interpretation are based either on
the loss-cone instability of trapped electrons producing electron cyclotron maser emission or on
upper-hybrid and Bernstein modes. An open question remains whether
or not spikes are signatures of particle accelerations episodes at a highly fragmented
energy release flare site. }
\item{Narrow Band Type III Bursts: Short ($\approx 1 s$) narrowband ($<200 MHz$) fast drifting ($>100 MHz/s$) bursts.
A number of this type of Bursts, on the SAO high resolution dynamic spectra,
exhibit a frequency drift turn over; as they drift towards lower frequencies and after
reaching a minimum frequency (\textit{turn over frequency}) they reverse direction towards
higher frequencies appearing as inverted U on the dynamic spectra. These we have marked as
narrowband type III(U) on figures \ref{FS1}, \ref{FS1B} \& \ref{FS1C}. 
Similar spectra (III(U), III(N)) were obtained in the microwave range by
\citet{Fu04}}.
\item{Narrow Band Slowly Drifting Bursts: They are similar to
Narrow Band Type III Bursts but with a drift rate $< 100 MHz/s$.}
\end{itemize}

\section{Summary and Discussion}
The ARTEMIS-IV radio-spectrograph, operating in the range of 650-20 MHz,
observed a number of complex events during the super-active period of period 14--20 January 2005;
the event on January the 17th was characterized by an extended, broadband type IV continuum
with rich fine structure.

We have examined the morphological characteristics of fine structure
elements embedded in the continuum;it, almost, matches the comprehensive Ondrejov Catalogue
\citep{Jiricka01,Jiricka02}. The latter, although it refers to the spectral range 0.8--2 GHz, seems to produce
similar fine structure with the metric range.

The high resolution (100 samples/sec) SAO recordings facilitated the spectral study
of the fine structures and permitted the recognition and classification of the
type III(U) \& III(J) subcategory of the Narrow Band Type III Bursts in the metric frequency
range; similar structures have been reported in the microwaves \citep{Fu04}.

The pulsating structures and fibers, although they cover the full observation interval, 
appear enhanced during the SXR rise phase and the two CME lift off where the major
magnetic restructuring takes place. The narrowband 
structures, on the other hand, are evenly distributed for the above mentioned 
duration; this indicates that small electron populations are 
accelerated even after the flare impulsive phase.

Two types of fine structures from the Ondrejov Catalogue were not detected in our recordings:
\begin{itemize}
\item{Continua: As the long duration pulsations accompanied by fibers were prevalent in the SAO spectra,
any possible appearance of Continua was, probably, suppressed within the pulsating background.}
\item{Lace Pattern: It is new type of fine structure
first reported by \citet{Jiricka01}; it is characterized by
rapid frequency variations, both positive and negative. It is a very rare structure with only
nine reported in the Ondrejov catalog out of a total of 989 structures.}
\end{itemize}

\ack{This work was supported in part by the Greek Secretariat for Research and Technology.
The authors would like to thank Prof. C. Caroubalos  (University of Athens)
for useful discussions and very helpful comments. They also acknowledge many useful suggestions
by the two unknown referees; these have significantly improved the quality original draft.}

\end{document}